\documentclass[12pt]{iopart}

\usepackage{graphicx}
\usepackage{hyperref}
\begin{document}

\title[Demonstration experiments using a table top mechanical Stirling refrigerator]{Demonstration experiments for solid state physics using a table top mechanical Stirling refrigerator}

\author{M R Osorio, A Palacio Morales\footnote{Present address: CEA Grenoble- INAC/SPSMS/IMAPEC 17, rue des Martyrs, 38054 Grenoble cedex 9, France}, J G Rodrigo, H Suderow and S Vieira}

\address{Laboratorio de Bajas Temperaturas, Departamento de F{\'i}sica de la Materia Condensada,
Instituto de Ciencia de Materiales Nicol{\'a}s Cabrera, Facultad de Ciencias, Universidad Aut{\'o}noma de Madrid,
E-28049 Madrid, Spain}

\ead{hermann.suderow@uam.es}

\begin{abstract}
Liquid free cryogenic devices are acquiring importance in basic science and engineering. But they can also lead to improvements in teaching low temperature an solid state physics to graduate students and specialists. Most of the devices are relatively expensive, but small sized equipment is slowly becoming available. Here, we have designed several simple experiments which can be performed using a small Stirling refrigerator. We discuss the measurement of the critical current and temperature of a bulk YBa$_2$Cu$_3$O$_{7-\delta}$ (YBCO) sample, the observation of the levitation of a magnet over a YBCO disk when cooled below the critical temperature and the observation of a phase transition using ac calorimetry. The equipment can be easily handled by students, and also used to teach the principles of liquid free cooling.
\end{abstract}

\submitto{\EJP}
\maketitle

\section{Introduction}
Cryogenic experiments for students are available in some universities for undergraduate students, but it must be admitted that they are not very common. We are often confronted in Madrid with students coming from other Universities with an important research activity at temperatures close to absolute zero, which have not been acquainted with basic cryogenic phenomena. Although the use of liquid nitrogen is very common, typical difficulties of handling cryogens and the need to be rather careful with heat leaks, cabling or other problems inherent to cryogenics are probable reasons for the lack of undergraduate student's work in cryogenic experiments. For instance, the observation of magnetic levitation of a high temperature superconductor in liquid nitrogen is of course widespread.  However, measuring its resistance or its critical current as a function of temperature is less common. Specific heat experiments showing phase transitions are even more rare. Recently, kits providing high temperature superconductors including wiring have been made available by several providers \cite{ColoradoSup}. On the other hand, dry cryogenic devices have entered the field of research laboratories, providing for cold environments without the use of liquid cryogens. In particular, Stirling refrigerators of small size, like the MicroStar K535 (RICOR, Israel \footnote{Further information and technical details can be found at the Ricor home page \cite{Ricor}.}) offer the possibility of refrigerating samples in a very simple manner, so they can be easily handled by students which, in turn, can acquire a valuable know-how concerning the functioning of this kind of devices. In addition to these advantages, the table top refrigerator allows the sample to be directly observed just by adding a glass lid onto the sample's holder.  Therefore, these simple cryogenic devices allow for easy, hands-on table-top operation with the additional pedagogical content of the Stirling refrigerator. We can expect that these devices will enter the range of student's laboratories. One single device can be used to make many experiments, thus making the realization of solid state physics experiments easier. 

In this paper, we suggest how a table top refrigerator can be used to perform different experiments at low temperatures. We show the demonstration of the first principles of high temperature superconductivity by observing the Meissner effect (levitation of a magnet due to magnetic field expulsion), and by measuring the $V$-$I$ and resistivity versus temperature curves of a YBa$_2$Cu$_3$O$_{7-\delta}$ (YBCO) disk. In addition, we used the MicroStar to find the temperature of the ferroelectric transition for a KH$_2$PO$_4$ (KDP) sample by means of the ac calorimetry method.

\section{Theory}

\subsection{Superconducting properties}
Superconducting cuprates have been extensively studied during the last decades \cite{Tinkham}, specially the hig-Tc material YBa$_{2}$Cu$_{3}$O$_{7-\delta}$ (YBCO), and the first two practices are intended to find both the critical temperature, $T_{\textrm c}$, and current, $I_{\textrm c}$, of a piece of this material. 

\subsubsection{Resistivity and critical temperature}

In order to find $T_{\textrm {c}}$, we propose to realize a direct measurement of the variation of the resistivity, $\rho$, as a function of temperature. As it is well known, $\rho$ decreases linearly with temperature, till $T_{\textrm {c}}$ is attained. The linear decrease of the resistance with temperature down to the lowest temperatures, has been associated to the peculiar metallic properties of these materials, and pertains, when it is found in high quality single crystals, to one of the debated issues in topical research in novel superconductors \cite{Mackenzie1996}. Hence, the resistivity decreases abruptly towards zero. Typically, in YBCO, we find around $T_{\textrm c}=90$ K, depending on sample's characteristics. This measurement allows discussing, for example, the role played by aging and inhomogeneities in the sample as well as the thermal fluctuations in the electron scattering. The importance of the former can be studied by inspecting the sharpness of the transition, expressed in terms of the width of the derivative of the resistance with respect to temperature at half maximum, that we denote as $\Delta T_{\textrm {cI}}$. This parameter depends strongly on the existence of different superconducting or normal phases, lack of oxygen, etc. Any region within a sample which behaves in a different way will have its own critical temperature, and so we will observe the result of the successive transitions of all of these small regions. On the other way, the thermal fluctuations provoke the rounding of the curve before the abrupt fall of the resistivity.

\subsubsection{Voltage-current curves} \label{sec:secVI} 

\begin{figure}
	\centering
		\includegraphics[width=0.5\textwidth]{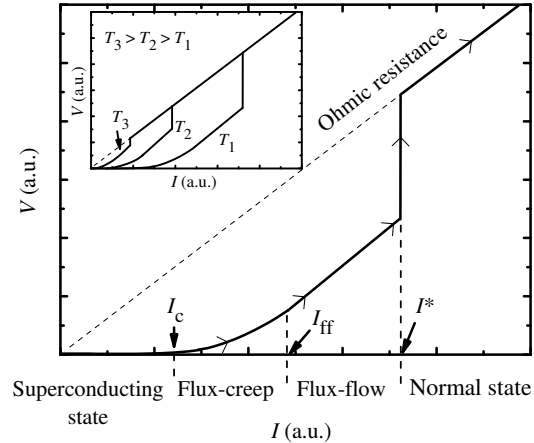}
	\caption{\footnotesize Characteristic $V$-$I$ curve for a YBCO sample. We can distinguish different regimes : superconducting state ($I<I_{c}$ i.e, $V=0$), flux-creep ($I>I_c$ and $V\propto I^n$), flux-flow ($I>I_{\textrm {ff}}$ and $V\propto (I-I_{\textrm {ff}}$)), and normal state ($I>I^*$). The insert shows how these curves change with increasing temperature ($T_i<T_{\textrm {c}}$, i=1, 2, 3,...). Once $T_{\textrm {c}}$ is attained, the resistance is ohmic for any current. See text for further explanations.}\label{fig:VItheoretical}
\end{figure}

Concerning the characteristic $V$-$I$ curves, they illustrate the different regimes of dissipation as well as the dependence of the critical current on sample's temperature\footnote{The critical current is also a function of the applied magnetic field whether it exists.}. \Fref{fig:VItheoretical} displays a description of a typical $V$-$I$ curve and the definition of different currents we will talk about in brief. For YBCO and similar superconductors, the magnetic field can penetrate inside the sample above a certain threshold, the lower critical field $H_{\textrm {c1}}$, but only in some small normal regions surrounded by supercurrent vortices deployed in a triangular array, whereas the rest of the material remains superconducting \cite{Tinkham}. The vortices may be pinned to potential wells associated to lattice defects (dislocations, impurities, ...) at sufficiently small values of applied current ($I<I_{\textrm c}$), so they do not move and the dissipation is zero. When the applied current is above a certain critical threshold ($I>I_{\textrm c}$), the Lorentz force per unit volume,

\begin{equation}
{\textbf F}={\textbf J}\times {\textbf B}
\label{Lorentz}
\end{equation}
${\textbf J}$ being the current density and ${\textbf B}$ the magnetic induction, is strong enough to move the vortices from potential wells, so they start to move and dissipation arises. This leads to two different regimes: flux-creep and flux-flow. The first one starts when the applied current is high enough to let temperature fluctuations remove vortices from their pinning centers \cite{Tinkham,Anderson1962,Feigelman1991,Soulen1994,Mints2005}. Flux-flow corresponds to the free drift of completely unpinned vortices under the force exerted by the applied current \cite{Tinkham,Prester1998}.
Both regimes can be normally distinguished by inspecting the characteristic curves: flux-creep exhibits a power-law dependence, i.e. $V\propto (I-I_{\textrm c})^n$ (with $n>1)$, and flux-flow follows $V\propto (I-I_{\textrm {ff}})$, where $I_{\textrm {ff}}$ is the onset current for this regime. Note that $I_{\textrm {ff}}>I_{\textrm c}>0$ always, so this linear behaviour is not ohmic. At sufficiently high currents, above the supercritical value $I^*$, the normal state is attained even if at that moment the temperature is well below $T_{\textrm c}$\footnote{See, for instance, references \cite{Vina2003,Maza2011}.}. The inset of \Fref{fig:VItheoretical} shows how these $V$-$I$ curves evolve as temperature is increased. Note that the nearer the temperature is to $T_{\textrm c}$, the closer is $I^*$ to $I_{\textrm c}$. 

The temperature increment can be due to the own sample's heating, because of dissipation. In that case, sample's temperature would increase as the $V$-$I$ curve is being traced. Feeding currents during a long time can provoke a noticeable heating in the metallic contacts. In that case, if the applied current is close to the critical value, this spurious heat can induce the early transition of the superconducting material. Nevertheless, the heat production is also a problem in the sample itself, once the critical current is overtaken. In this scenario, a {\it long} time can be several seconds for some low critical current density samples or even ms or $\mu$s for thin-film microbridges. The importance of these aspects and the need of a fast acquisition system which applies short current pulses is remarked, for instance, in references \cite{Gonzalez2003,Ruibal2007,Osorio2006}. 

\subsubsection{Meissner effect and levitation}

The levitation observed when approaching superconducting samples and magnets has been extensively studied \cite{Brandt1989,Chen1990,Chen1992}. The Meissner effect i.e., the expulsion of the magnetic field from the superconductor, is a consequence of the existence of supercurrents which cancel the magnetic field in any point of the superconductor. 
YBCO is a type-II superconductor, where vortices get trapped at the pinning centers (inhomogeneities, lattice defects, etc). 
Trapped flux lines allow for the establishment of an interaction between magnet and superconductor, and thus to magnetic levitation \cite{Brandt1989,Saito2010}, as can be observed in many videos available in the Internet \cite{Dresden,UAM1,UAM2}. Practical applications include levitation trains and magnetic bearings \cite{Gabrys1992,Ma2000,Okano2003,Nagashima2009}, whose aim is taking advantage of the lack of friction between moving parts made from superconductors and magnets.

\subsection{Heat capacity measurements: ac calorimetry method}

The ac calorimetry method \cite{Sullivan1968,Gmelin1997,Devoille2002} allows measuring changes in the heat capacity quite accurately. It is therefore a very good method to measure the critical temperature of phase transitions in different materials.
The ac calorimetry method is easy to implement, and very illustrative. The jump in the specific heat at the superconducting transition is very difficult to observe, because the electronic specific heat term is hidden inside a large phonon background \cite{Inderhees1987}. It is therefore more convenient, in an undergraduate  student's lab, to measure other more handy transitions, such as the ferroelectric transition.

When applying a sinusoidal heating to a sample, $\dot{\mathcal Q}= \dot{\mathcal Q_0}(\cos{\frac{1}{2}\omega t})^2$, its equilibrium temperature has a term which is inversely proportional to its heat capacity. This term can be greatly simplified by taking into account some reasonable assumptions, and the heat capacity can be then directly estimated.

\begin{figure}
	\centering
		\includegraphics[width=0.5\textwidth]{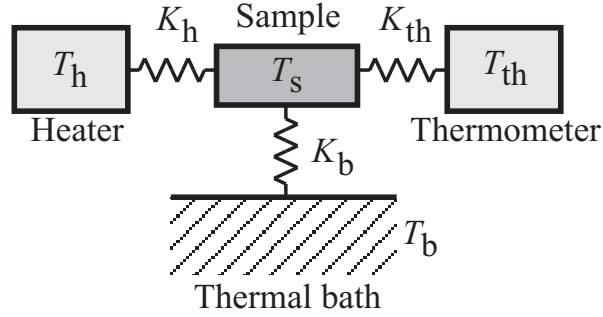}
	\caption{\footnotesize Scheme of the interconnections between the sample and the thermal bath, the heater and the thermometer.}\label{fig:Calorimetry}
\end{figure}

In order to go into this method's details, let us consider the scheme of \Fref{fig:Calorimetry}, where a sample (with heat capacity $C_{\textrm s}$) is coupled to a bath, a thermometer ($C_{\textrm {th}}$) and a heater ($C_{\textrm h}$) through thermal conductances $K_{\textrm {b}}$, $K_{\textrm {th}}$ and $K_{\textrm {h}}$, respectively. These elements are at temperatures $T_{\textrm s}$, $T_{\textrm b}$, $T_{\textrm {th}}$ and $T_{\textrm h}$, respectively. The equations governing the whole system are then:

\begin{eqnarray}
C_{\textrm {h}}\frac{\partial T_{\textrm {h}}}{\partial t}&=\frac{\partial \mathcal Q_0}{\partial t}\left(\cos{\frac{1}{2}\omega t}\right)^2-K_{\textrm {h}}(T_{\textrm {h}}-T_{\textrm {s}}) \nonumber \\ 
C_{\textrm {s}}\frac{\partial T_{\textrm {s}}}{\partial t}&=K_{\textrm {h}}(T_{\textrm {h}}-T_{\textrm {s}}) -K_{\textrm {b}}(T_{\textrm {s}}-T_{\textrm {b}})-K_{\textrm {th}}(T_{\textrm {s}}-T_{\textrm {th}})  \\
C_{\textrm {th}}\frac{\partial T_{\textrm {th}}}{\partial t}&=K_{\textrm {th}}(T_{\textrm {s}}-T_{\textrm {th}}). \nonumber 
\label{Thermal-eq}
\end{eqnarray}

The detailed resolution of this array of equations can be found elsewhere \cite{Sullivan1968,Gmelin1997}. It can be assumed that, for small enough temperature variations, both the heat capacities and the thermal conductivities are constant as the temperature changes, and that the thermal bath remains stable. The solution for the stationary temperature yields a constant term depending on $K_{\textrm {b}}$ and an oscillatory term, $T_{\textrm {ac}}$, which is inversely proportional to the heat capacity of the sample. This dependence can be expressed as simply as:

\begin{equation}
C_{\textrm s}=\frac{\dot{\mathcal Q_{\textrm {0}}}}{2\omega T_{\textrm {ac}}}.
\label{Heat-capacity}
\end{equation} 

In order to arrive to \Eref{Heat-capacity}, several assumptions have to be made:

\begin{itemize}
\item The heat capacities of the heater, the thermometer and any existing supporting piece are negligible compared to that of the sample (i.e., $C_{\textrm  s}\gg C_{\textrm  {th}}, C_{\textrm  h}, ...$).
\item The condition $\omega^2(\tau_{\textrm {th}}^2+\tau_{\textrm {h}}^2)\ll 1$ is satisfied, $\tau_{\textrm {th}}=C_{\textrm {th}}(K_{\textrm {th}})^{-1}$ and $\tau_{\textrm {h}}=C_{\textrm {h}}(K_{\textrm {h}})^{-1}$ being the relaxation times of both the thermometer and the heater. This means that the heater and the thermometer come to equilibrium in a time interval much shorter than the inverse of the excitation frequency.
\item The frequency is much larger than the inverse of the sample-to-bath relaxation time.
\end{itemize}

To implement this experiment, some conditions regarding \Eref{Heat-capacity} need to be fulfilled, as we are going to see immediately.
Let us define the external relaxation time, $\tau_{\textrm {ext}}$, as the time needed to reach the thermal equilibrium between sample and bath. On the other hand, the internal relaxation time, $\tau_{\textrm {int}}$, will be the sum of the relaxation times between thermometer and sample as well as between heater and sample. Finally, let us take the own sample's relaxation time, $\tau_{\textrm {s}}$, as approximately equal to $\tau_{\textrm {int}}$. The set of conditions to be satisfied if we want to apply \Eref{Heat-capacity} is \cite{Sullivan1968,Gmelin1997}:

\begin{equation}
\tau_{\textrm {ext}}>100\tau_{\textrm {int}},
\label{tauext}
\end{equation}

\begin{equation}
\frac{\tau_{\textrm {ext}}}{10}>\frac{1}{\omega}>10\tau_{\textrm {int}}.
\label{tauext-freq}
\end{equation}
\Eref{tauext} indicates that the relaxation time between sample and bath must be much longer than the internal relaxation time of the sample. \Eref{tauext-freq} states that the period of the oscillating signal must be also much longer than $\tau_{\textrm {int}}$. Note that this modulation method demands at least a weak heat loss to the bath. A completely adiabatic cell, with $\tau_{\textrm {ext}}\to\infty$, would never work.

All the relaxation times involved in this practice can be calculated from \Eref{Relaxation-time}:
\begin{equation}
\tau=\frac{C}{K}=\frac{C}{\kappa}\frac{\ell}{S},
\label{Relaxation-time}
\end{equation}
where $\ell$ and $S$ are characteristic dimensions (in the case of the sample, its thickness and base area) and $\kappa$ the thermal conductivity. To be sure that we are in the scenario of applicability of \Eref{Heat-capacity}, the excitation frequency and every piece on our set-up have to be carefully chosen (regarding its thermal conductivity and dimensions) so the relaxation times, $\tau_{\textrm {ext}}$ and $\tau_{\textrm {int}}$, take the right values. Thinking about a suitable and simple way to achieve this is a very formative matter to be addressed to the students. It can be considered as a good practical exercise on the role played by the thermal conductivity and the heat capacity of materials, realizing how the relaxation times vary as we use copper or stainless steel, for instance. In addition, ancillary experiments can be devoted to the practical finding of these magnitudes \cite{Ortuno2011,Lewowski2001}. In section \ref{sec:ac} we propose simple solutions for a suitable set-up.

\subsection{Stirling refrigerator characteristics}

\begin{figure}
	\centering
		\includegraphics[width=0.5\textwidth]{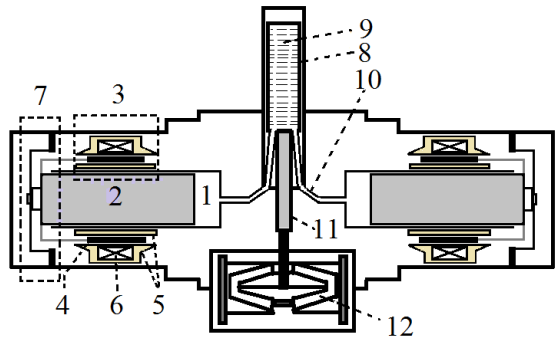}
	\caption{\footnotesize Scheme of the MicroStar Stirling refrigerator. See text for details.}\label{fig:Scheme}
\end{figure}

The MicroStar Stirling refrigerator is comprised of two pistons symmetrically located at the sides of a combined expander-regenerator, as depicted in \Fref{fig:Scheme} (see \cite{Riabzev2009} for a more detailed explanation). Both pistons (2) reciprocate inside the cylinders (1), so the gas is forced through two channels (10) which feed an expander-regenerator assembly ((8) and (9)). Its driving plunger (11) is tied to a spring compensation system (12), allowing negligible spurious vibrations. Two magnetic motors (3) produce the reciprocation of the pistons. They are made up from a permanent magnet (4) that is located in the air gap of a magnetic core (5), in which an alternating magnetic field is created by feeding a primary coil, (6). As the polarity of the field changes according to the frequency of the current, the permanent magnet reciprocates in the gap as it is attracted and repelled in each half cycle. The movement is transmitted to the pistons by means of planar springs (7).

\begin{figure}
	\centering
		\includegraphics[width=0.5\textwidth]{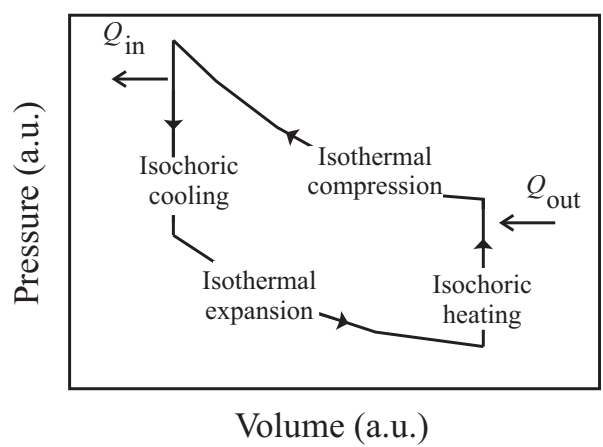}
	\caption{\footnotesize Ideal Stirling cycle. The cooling power is produced during the isothermal expansion.}\label{fig:Stirling}
\end{figure}

The difference established between the movement of the pair of pistons and the expander-regenerator (about $90^\circ$) allows a Stirling cycle like that of \Fref{fig:Stirling} to be set inside the gas cavity and so yields cooling power at the end of the expander-regenerator. Let us remember that the Stirling cycle is made up from two isothermal and two isochoric process and that cooling power arises from the isothermal expansion. During this step, the temperature of the gas remains nearly constant as heat is absorbed by the gas from the cold tip wall which, in turn, is chilled a bit more on every cycle \cite{Stirling}.

The device can be controlled by using a computer (a commercial program is available) as well as by a manual controller. The temperature inside the cold cavity can be monitored by using the own thermometer of the MicroStar, so any additional wiring is not necessary apart from the current and voltage cables arising from the sample, if needed.
The cooling power decreases with temperature ($6$ W at $65$ K, for instance).

Amongst the many advantages of this device, it is to be remarked its low mechanical noise level, as well as its capacity to reach about $55$ K just in a few minutes. But the more interesting point is the easiness of use and that it is really a table-top device, without additional big compressors, and that it makes a nice, extendable, hands-on machine for practicing.

\section{Experiments}

\subsection{Resistivity and $V$-$I$ characteristic curves}

\begin{figure}
	\centering
		\includegraphics[width=0.3\textwidth]{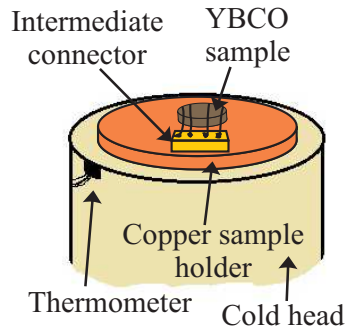}
	\caption{\footnotesize Top of the expander-regenerator, where the sample holder was attached by four screws (not visible in this scheme). An intermediate connector allowed changing the sample easily.}\label{fig:Sample-holder}
\end{figure}

We propose three types of experiments to be carried out with superconductors; the first two ones, resistivity vs temperature and $V$-$I$ characteristic curves, can be realized with the same set-up.
We attached four thin copper ribbons about $1$ mm wide to a bulk YBCO disk of $12$ mm in diameter and $2.5$ mm in thickness by using a silver conductive epoxy (Silver Loaded Epoxy Adhesive, reference number RS 186-3616, supplied by RS Components). After that, the sample was kept inside a furnace at about $70^{\circ}$C during an hour, in order to harden the epoxy.
Then, the sample was fixed with Apiezon grease to a cooper sample-holder of about $2$ mm in thickness, which was attached to the cold head of the table top refrigerator by using suitable screws (see \Fref{fig:Sample-holder}).

Once these steps were accomplished, the cavity was sealed and pumped with a rotary pump. A current of $26$ mA was fed to the sample and a voltmeter was used to acquire the sample's voltage.
Then, the temperature was decreased while the variation of the voltage was monitorized. In general, a thermoelectric voltage is to be expected in this type of measurements, due to differences in the temperature of the contact pads. If we suppose that they are at two different temperatures, $T_1$ and $T_2$, the thermoelectric voltage will depend on the Seebeck coefficients or thermopowers of the different materials that make up the contact pads \cite{Guenault1978} i.e., the copper ribbons and the silver conductive epoxy. Let us call them $S_{\textrm {A}}$ and $S_{\textrm {B}}$. Hence, the thermoelectric voltage would be given by \Eref{Seebeck} 

\begin{equation}
V = \displaystyle\int_{T_1}^{T_2} \left(S_{\textrm A}(T)-S_{\textrm B}(T)\right)\, dT.  \label{Seebeck}
\end{equation}

\begin{figure}
	\centering
		\includegraphics[width=0.5\textwidth]{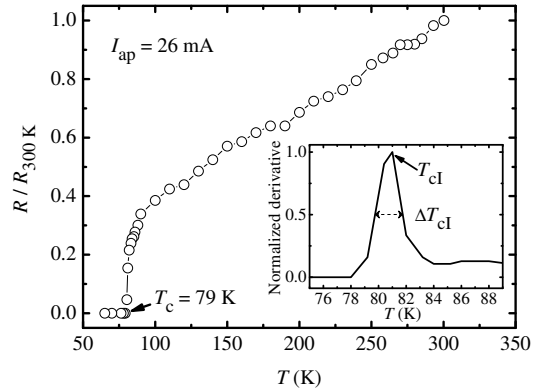}
	\caption{\footnotesize Resistance vs temperature curve obtained when cooling down the sample. $R$ is normalized to its value at $300$ K. The inset shows the derivative of this curve normalized to the maximum value, and the definition of $T_{\textrm {cI}}$ and $\Delta T_{\textrm {cI}}$. See text for details.}\label{fig:RT-Curve}
\end{figure}

To remove its contribution it is necessary to reverse the polarity in the current source, so we acquired both $V(I^+)$ and $V(I^-)$. As the thermoelectric voltage is the same no matter the polarity, we can get the right voltage by using \Eref{Thermal-V}:

\begin{equation}
V = \frac{V(I^+)-V(I^-)}{2}. \label{Thermal-V}
\end{equation}
\Fref{fig:RT-Curve} displays the resistance vs temperature curve for our YBCO disk and the aforementioned applied current (the resistance is normalized to its value at $300$ K). We found a critical temperature $T_{\textrm {c0}}=79$ K. This low value is not surprising, as the applied current was high enough to reduce the critical temperature (for bulk YBCO $T_{\textrm {c}}$ uses to be about about $90$ K). Later on, we will comment on the temperature dependence of the critical current. 
The inset shows the derivative of the resistance with respect to temperature, normalized to the maximum value. We can see that the inflection point of the derivative, denoted as $T_{\textrm {cI}}$, is at $81$ K, and that $\Delta T_{\textrm {cI}}=2$ K, which is quite a large value. The origin of the transition width can be discussed further in connection with the critical current measurements shown below.

Due to the fast action of the Stirling refrigerator ($70$ K can be attained in less than $20$ min), a thermal gradient is normally established between the thermometer and the sample. Hence, it is important to let the system thermalize before data are acquired (about $10$ min, at least). The refrigerator allows setting a specific temperature value, so the cooling can be realized quite slowly. Although for high temperatures (above $100$ K) there is a remarkable oscillation due to the feed-back system, it is to be noticed that the lower the temperature, the better the stabilization the device can yield.

\begin{figure}
	\centering
		\includegraphics[width=0.5\textwidth]{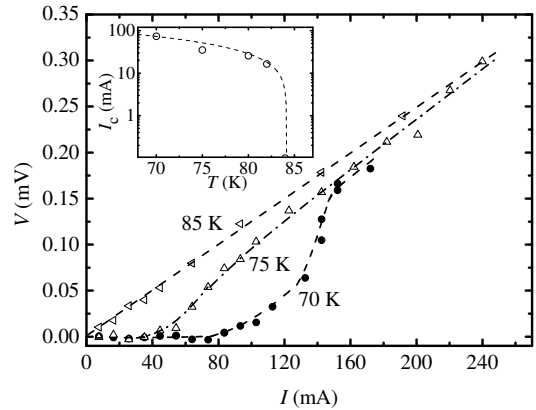}
	\caption{\footnotesize Some characteristic curves of the YBCO sample at several temperatures. Above $T\approx84$ K the behaviour is totally ohmic. The inset shows the $I_{\textrm {c}}$ vs $T$ curve in logarithmic scale. Dashed lines are just guides to the eye. These curves can be now compared with the theoretical ones of \Fref{fig:VItheoretical}.}\label{fig:V-I}
\end{figure}

\Fref{fig:V-I} shows several characteristic curves obtained at $70$ K, $75$ K and $85$ K (one of the advantages of this device is its capability to go well down $77$ K, which is the limit we could afford by using liquid nitrogen). Dashed lines are just a guide to the eye. The inset displays the dependence of the critical current on the temperature in logarithmic scale, which stress the sudden fall in the vicinity of the transition. 
It is interesting to note that we obtained these curves without a fast acquisition system. As a consequence, our measurements are somewhat affected by spurious heat generation, as we commented in the last paragraph of section \ref{sec:secVI}. This can be observed in the smooth transition to the ohmic state for the lowest temperature ($T=70$ K), where flux-flow regime cannot be clearly discerned. It can be noticed, on the contrary, the rounding of the curve just above the transition, corresponding to flux-creep phenomenon. For $T=75$ K the flux-creep regime leads to a linear but non-ohmic behaviour, hence a free vortex movement or flux-flow, before the normal state is attained. In this case the critical current is lower, and then the dissipation a bit less important. For $T=85$ K we observe the normal state even for the lowest applied currents. 

Despite we measured these curves with a slow and quite cheap acquisition system (a conventional voltmeter and a basic current source, made at the workshops of the University, the Segainvex, whose current value can be driven by an external voltage), we can distinguish the different dissipation regimes and the formative aim was perfectly satisfied.

\subsection{Levitation of a magnet over a superconducting disk}

\begin{figure}
	\centering
		\includegraphics[width=0.5\textwidth]{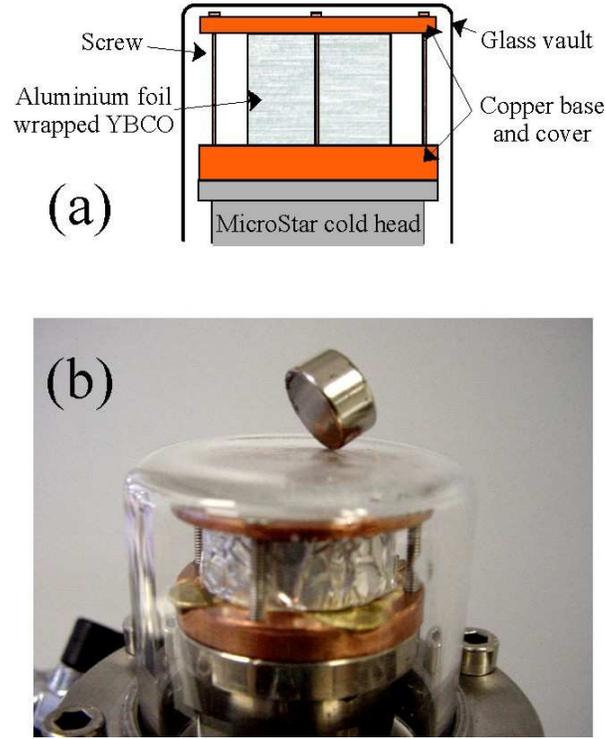}
	\caption{\footnotesize (a) Set-up for the observation of the Meissner effect. (b) Levitation of a magnet over the cold superconducting disk.}\label{fig:Levitation}
\end{figure}

In order to make possible the observation of the levitation, a glass vault was built and located on top of the cold head, in such a manner that the superconducting YBCO disk ($2$ cm in diameter and about $1$ cm in thickness), mechanically attached to the cold tip, remained at a very small distance from the top of the vault (see \Fref{fig:Levitation}(a)). Two copper plates surrounded the YBCO disk. That at the bottom acted like a thermal anchor, whereas the one located on top of the sample assured the heat extraction from that part, thus reducing the possible thermal gradients.

It is important that the screws are not in contact with the glass vault, or a thermal gradient could appear in the sample, the hottest point being at ambient temperature. A visual inspection is normally enough to be sure they are far enough from the glass. In case of contact, condensed water would cumulate on it. 
As a precaution, to limit the incidence of radiation (a remarkable aspect, the YBCO disk being black), the superconductor was wrapped in a thin aluminium foil. This effect allows discussing radiation losses at low temperatures, an important aspect of cryogenic devices. We can make a very simple estimation of the heat absorbed by radiation with and without the aluminium foil. Let us assume that heat is radiated just from the upper face of the YBCO disk and that it has the same area, $A$, as the glass vault above it (we make the approximation of plane parallel surfaces). The disk and the glass are at equilibrium temperatures $T_2$ and $T_1$, respectively, and they are characterized by emissivities $\epsilon_2$ and $\epsilon_1$, respectively\footnote{Emissivity is the ratio of energy radiated by a material to energy radiated by a black body at the same temperature. It is close to zero for a very good reflector and is equal to $1$ for a black body.}. Then, the rate of heat transfer by radiation is given by (chapter 4 of reference \cite{White}):

\begin{equation}
\dot{\mathcal Q}=\sigma A(T_1^4-T_2^4)\frac{\epsilon_1\epsilon_2}{\epsilon_1+\epsilon_2-\epsilon_1\epsilon_2},
\label{Radiation}
\end{equation}
$\sigma=5.67\times 10^{-8}$ Wm$^{-2}$K$^{-4}$ being the Stefan-Boltzmann constant. To simplify, let us suppose that the superconducting disk is wrapped in a clean polished foil of aluminium. Then, $\epsilon_2=0.02$ at about $80$ K, and for the glass lid $\epsilon_1=0.9$ at $300$ K \cite{White}. In our experiment, $T_2\approx 80$ K, $T_1\approx 300$ K, and the area of both surfaces is $A=3$ cm$^2$. Hence, the rate of heat transfer by radiation is $3$ mW. On the contrary, if the YBCO disk is left unwrapped $\epsilon_2\to 1$, and $\dot{\mathcal Q}=130$ mW, which is by far a higher heat leak. 

Once all these aspects were taken into account, a small magnet made from a neodymium alloy (NdFeB from Supermagnete) was located on top of the glass vault. 
On cooling down, as the local temperature went below the critical value, $T_{\textrm {c}}$, the magnet began to separate from the glass cover. \Fref{fig:Levitation}(b) shows an image of the levitation of one of these magnets.

\subsection{Observation of the ferroelectric transition of a KDP sample}\label{sec:ac}

As a final example of the possibilities of the table top refrigerator, we are going to describe now the experiment intended to obtain the ferroelectric transition temperature for a Potassium Dihydrogen Phosphate sample (KH$_2$PO$_4$). For that, we used the ac calorimetry procedure described above.

\begin{figure}
	\centering
		\includegraphics[width=0.4\textwidth]{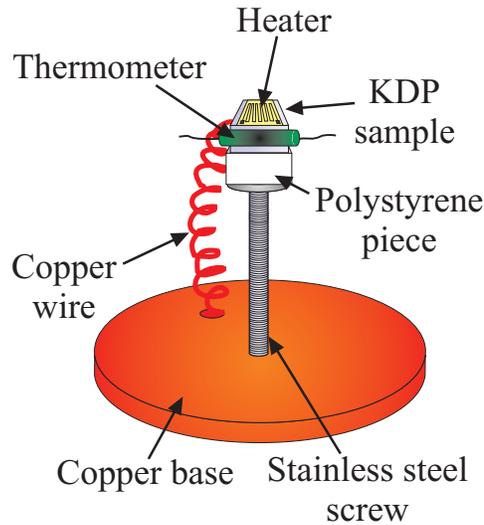}
	\caption{\footnotesize Calorimetric cell used to realize the heat capacity measurements. The radiation shield and the wires are not included.}\label{fig:Scheme-KDP}
\end{figure}

The set-up is depicted in \Fref{fig:Scheme-KDP}. The KDP element we used is $5.90$ mm long, $5.75$ mm wide and $2.50$ mm thick, and it weighs $163.1$ mg. Taking into consideration that the heat capacity of KDP is $C_{\textrm {KDP}}\simeq 1$ Jg$^{-1}$K$^{-1}$ \cite{Kumzerov2011} and its thermal conductivity is $\kappa_{\textrm {KDP}}\simeq 1.9\times10^{-2}$ Wcm$^{-1}$K$^{-1}$ (at about $120$ K \cite{Yoreo1985,Fengqi1994}) the internal relaxation time is $\tau_{\textrm {int}}\approx\tau_{\textrm {s}}=C_{\textrm {KDP}}(K_{\textrm {KDP}})^{-1}=4.65$ s. 

The sample was glued with a suitable varnish onto a small piece of expanded polystyrene which approximately the same lateral dimensions, and about $2$ mm in thickness. Due to its intrinsic properties\footnote{A technical description of polystyrene can be found, for instance, at \cite{Polystyrene}.} ($\kappa_{\textrm {pol}}\approx 0.03$ Wm$^{-1}$K$^{-1}$ and $C_{\textrm {pol}}\approx 1.1$ KJKg$^{-1}$K$^{-1}$ at about $280$ K, and these magnitudes decrease when reducing the working temperature) and very low mass ($4.2$ mg), it acted as an effective thermal barrier and its contribution to the heat capacity was negligible ($0.005$ JK$^{-1}$ for the polystyrene and $0.163$ JK$^{-1}$ for the sample). 

By using the same varnish, we attached a transducer-class strain gage bondable resistor (Micro-Measurements, USA) on top of the sample, as well as a carbon-ceramic sensor (TMI Cryogenic) at one side\footnote{Their characteristics are resumed at \cite{MM} and \cite{Temati}, respectively.}. The transducer was used as a heater, connected to the source through copper wires, and the carbon-ceramic sensor as a thermometer, fed with a $10\;\mu$A current by using manganine wires, in order to avoid heat loses. The heat capacities of both elements were considered as negligible.

This assembly was finally glued to a stainless steel screw, attached to a copper base that was in turn fixed to the cold tip of the refrigerator. The thermal contact between the cell and the cold tip was deliberately poor. Then, to allow a controlled thermal leak to the bath, we glued a spirally wounded copper wire to the KDP sample as well as to the base. The whole cell was surrounded by a copper wall and a cover in good contact with the cold tip of the table top refrigerator. They played the role of radiation shields to prevent that any heat coming from the room temperature walls beside could alter the sample's temperature.

With this configuration, the time needed to attain the thermal equilibrium between the sample and the bath is much longer than the internal relaxation time. For convenience, the copper wire used to link the cell to the heat sink (i.e. the bath) was $6$ cm in length and $0.2$ mm in diameter.
Taking $\kappa_{\textrm {Cu}}\approx500$ Wm$^{-1}$K$^{-1}$ as the thermal conductivity of copper within the interval of interest in our measurement (i.e, $100-130$ K)\footnote{At these temperatures both the contributions due to phonons and electrons are of relevance for the thermal conductivity. This value was obtained from chapter 11 of reference \cite{White}. Other cryogenic properties of copper and its alloys can be consulted at \cite{Copper}.}, the ``external'' relaxation time is $\tau_{\textrm {ext}}=C_{\textrm {KDP}}(K_{\textrm {Cu}})^{-1}=1387$ s (here, $\ell$ and $S$ are the length and the cross-section area of the copper wire).

Therefore, the conditions of \Eref{tauext} and \Eref{tauext-freq} are fulfilled whether it is selected a frequency in the range $0.007<\omega<0.021$, and we decided to feed current at $\omega=0.01$ Hz (a shorter $\tau_{\textrm {ext}}$ would further reduce the frequency window). 

\begin{figure}
	\centering
		\includegraphics[width=0.5\textwidth]{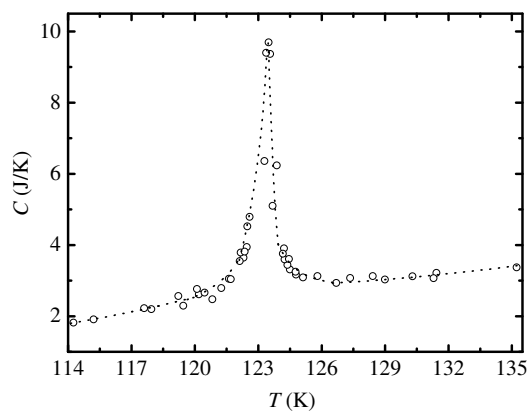}
	\caption{\footnotesize Heat capacity versus temperature for a KDP sample. The peak corresponds to the ferroelectric transition. The dotted line is just a guide for the eye.}\label{fig:C-KDP}
\end{figure}

The experimental procedure was as follows: first, we set a temperature about $80$ K in the cold tip. Once the cell had attained its equilibrium temperature (somewhat above that of the cold tip), we injected a current to the heater and we waited till the thermal equilibrium was reached in the sample (i.e., the temperature measured by the carbon-ceramic sensor was stable, except for the ac oscillation). We repeated these steps while progressively increasing the current fed to the heater, and till a temperature about $135$ K. \Fref{fig:C-KDP} shows the heat capacity versus temperature curve obtained during several runs.  The sudden change observed at around $T=123$ K corresponds to the ferroelectric transition \cite{Kumzerov2011,Fengqi1994,Stephenson1944}. The accuracy in the determination of the critical temperature strongly depends on the quality of the shield against radiation. Not only the copper barrier is necessary, but also a high vacuum inside the chamber. Poor vacuum, above $10^{-4}$ mbar, leads to an extra heat exchange between the sample and the non-evacuated gas around the cell. In this case, it can be expected a noticeable rounding of the peak and an apparent change in the transition temperature, maybe due to an inhomogeneous temperature distribution in the sample (i.e. the surface could be at a lower temperature than the inner parts, due to convective heat exchange with the remaining gas).

\section{Conclusions}

We have presented a brief description of some possible experiments which can be carried out with a miniature Stirling refrigerator (MicroStar, RICOR). We have demonstrated several experiments which allow discussing important educational aspects of solid state physics: (i) thermal and electrical resistance of metals, (ii) critical temperature of superconductors, (ii) flux expulsion, (iii) presence of vortices and its motion, (iv) detection of phase transitions through its effect in the specific heat. The additional advantage of reviewing the Stirling method, already studied in thermodynamics, on a table top apparatus which allows reaching temperatures down to 50 K, is an interesting plus for a student's laboratory.

\ack
The Laboratorio de Bajas Temperaturas is associated with the ICMM of the CSIC. This work
was supported by the Spanish MICINN (Consolider Ingenio Molecular Nanoscience CSD2007-
00010 program and FIS2008-00454 and by ACI-2009-0905), by the Comunidad de Madrid
through the program Nanobiomagnet (S2009/MAT1726) and by the NES program of the ESF.

\end{document}